\documentclass[runningheads]{llncs}
\usepackage{graphicx}
\usepackage{xcolor}
\usepackage{amssymb}
\usepackage{multirow}
\usepackage{booktabs}
\usepackage{makecell}

\begin{document}
\title{3D Reasoning for Unsupervised Anomaly Detection in Pediatric WbMRI}
%
\author{Alex Chang\inst{1, 2, 3} \and
Vinith Suriyakumar\inst{1, 2,3} \and Abhishek Moturu\inst{1,2, 3} \and James Tu\inst{3} \and Nipaporn Tewattanarat\inst{2} \and Sayali Joshi\inst{2} \and Andrea Doria\inst{2} \and Anna
Goldenberg\inst{1, 2, 3}}

\authorrunning{Chang A et al.}
%
\institute{University of Toronto, Toronto, Canada \and
The Hospital for Sick Children, Toronto, Canada \and
Vector Institute, Toronto, Canada}
\maketitle              
\begin{abstract}

Modern deep unsupervised learning methods have shown great promise for detecting diseases across a variety of medical imaging modalities. While previous generative modeling approaches successfully perform anomaly detection by learning the distribution of healthy 2D image slices, they process such slices independently and ignore the fact that they are correlated, all being sampled from a 3D volume. We show that incorporating the 3D context and processing whole-body MRI volumes is beneficial to distinguishing anomalies from their benign counterparts. In our work, we introduce a multi-channel sliding window generative model to perform lesion detection in whole-body MRI (wbMRI). Our experiments demonstrate that our proposed method significantly outperforms processing individual images in isolation and our ablations clearly show the importance of 3D reasoning. Moreover, our work also shows that it is beneficial to include additional patient-specific features to further improve anomaly detection in pediatric scans.

\end{abstract}

\section{Introduction}

Deep learning has shown promise in detecting cancerous lesions in breast~\cite{wang2016deep}, prostate~\cite{schelb2019classification} and lungs ~\cite{rossetto2017deep}. In earlier problem settings, there were too few training examples as well as insufficient number of disease labels preventing supervised methods from achieving good performance. Recently, unsupervised deep learning methods have started to show promise in these challenging settings by reformulating the problem as anomaly detection, using only the disease-free images during training and diseased images for testing ~\cite{schlegl2019f}. The use of these methods is still relatively unexplored in pediatric cancer detection setting. In pediatric cancer screening, especially in  diseases of cancer predisposition, which are intrinsically rare, whole-body MRI (wbMRI) is an essential part of the screening protocol~\cite{greer2017pediatric}. The goal is to detect lesions as early as possible, which is a difficult task for radiologists because of the small size of nodules and professional inexperience working with this modality. Unsupervised deep learning for anomaly detection is yet to be applied in this setting.

Existing techniques use variants of generative adversarial networks (GANs) \cite{goodfellow2014generative} and variational autoencoders (VAEs) \cite{kingma2019introduction} to model the distribution of non-cancerous patients and doing out-of-distribution detection to highlight anomalous cases, i.e. those with cancer ~\cite{schlegl2019f}. Most studies applying these techniques to MRIs have focused on single body parts such as the brain~\cite{mohsen2018classification}. These settings are less difficult than the wbMRI setting because all of the images are well registered and slices look similar throughout the volume. wbMRIs present challenges such as poor registration between patient volumes, large heterogeneity between the slices in the MRI volume, and lower signal-to-noise ratio. These issues are exacerbated in the pediatric setting because patients at various stages of growth and development look differently on this and other imaging modalities. Due to these problems, traditional registration is not applicable to the pediatric setting. We overcome these issues by building on unsupervised methods to leverage 3D information, and patient specific features.

We propose using a conditional VAE~\cite{sohn2015learning} with a sliding window approach to overcome these challenges for cancer detection in pediatric wbMRI (Figure~\ref{fig:overview}). While prior work has previously dealt with 3D inputs by either using computationally expensive 3D convolutions or by sampling individual slices as 2D inputs, our model accepts consecutive slices as multi-channel inputs and is conditioned on the slice and window coordinates to facilitate 3D reasoning. Moreover, we condition the model on patient features to disentangle the learned latent space to more accurately model variations in different regions of the body such as the brain, chest and abdomen. The contributions of our study are four-fold:



\begin{enumerate}
    \item We show that processing multiple consecutive slices of 3D volumes significantly improves unsupervised anomaly detection.
    \item We learn the distribution of large healthy body regions using a sliding window approach by informing the model of the window location.
    \item We show that conditioning on patient meta-features can lead to substantial improvements in anomaly detection. 
    \item We are the first to perform anomaly detection for pediatric wbMRIs which are more challenging than other settings due to difficulties in registration and images involving multiple body regions. 
\end{enumerate}

Overall, we demonstrate multiple techniques towards more accurate anomaly detection for cancer patients and we propose a method capable of processing challenging whole-body MRI scans. 


\section{Related Work}

\subsection{Whole-Body MRIs in Pediatric Cancer Screening Protocols}

Whole-body magnetic resonance imaging (wbMRI) is an essential part of well-established
cancer screening protocols~\cite{villani2016biochemical}. These protocols have shown improvements in early detection of cancer for both adult~\cite{attariwala2013whole} and pediatric~\cite{greer2017pediatric} patients. 

\subsection{Deep Learning for Cancer Detection}
Both supervised and unsupervised deep learning approaches have demonstrated potential for improving cancer detection in a number of settings. Supervised learning methods such as convolutional neural networks have been used to improve cancer detection across imaging modalities such as MRIs~\cite{hu2020deep}, X-rays~\cite{ausawalaithong2018automatic}, and CTs~\cite{ozdemir20193d}. Different cancers where this has been demonstrated include: breast cancer~\cite{wang2016deep}, prostate cancer~\cite{nagpal2019development}, brain cancer~\cite{mohsen2018classification}, and lung cancer~\cite{rossetto2017deep}. Specifically, for whole-body MRIs, supervised deep learning has shown promise in screening adult patients~\cite{lavdas2019machine}. To the best of our knowledge, these methods have not been applied to pediatric wbMRI cancer screening, where due to smaller sample sizes and potentially even higher class imbalance these methods are not likely to succeed.

Unsupervised deep learning techniques for cancer detection have focused on formulating the task as anomaly detection. This approach has shown superior performance to supervised deep learning techniques when the proportion of disease labels is low~\cite{chang2020using}. Unsupervised deep learning has also been used as a feature extractor to improve cancer detection. One example of this is in improving breast cancer detection~\cite{zhang2018integrating}. Initial work has shown the promise of generating single pediatric wbMRI slices and anomaly detection~\cite{chang2020using} but no work to our knowledge takes the full complexity into account successfully.

\subsection{Unsupervised Anomaly Detection with Generative Models}

Variations of VAEs, GANs, and combinations of the two have been proposed as solutions to anomaly detection. In cancer detection, these methods aim to model the distribution of patients without cancer and then perform out-of-distribution detection to highlight image regions with cancer. First, AnoGAN has demonstrated potential in highlighting anomalous regions on optical CT scans~\cite{schlegl2019f}. GANomaly\cite{akcay2018ganomaly} and f-AnoGAN\cite{Schlegl2019fAnoGANFU} improve this framework by removing the costly optimization steps required for testing AnoGAN. However, these methods are restricted to detection in low-resolution patches. AnoVAEGAN performed cancer segmentation in high-resolution brain MRI using a VAE-GAN hybrid with spatial latent space \cite{baur2018}. Although they performed cancer detection in volumetric brain MRI, an important drawback from their method is that they process each slice as an independent sample, resulting in 2D modeling which do not capture 3D context; only a postprocessing step is applied afterwards with the combined slices. Finally, Viana et al \cite{viana2020unsupervised} propose using a model based on f-AnoGAN which successfully handles volumetric brain MRI but requires computationally expensive 3D convolution operations.
\begin{figure}[t]

\includegraphics[width=\textwidth]{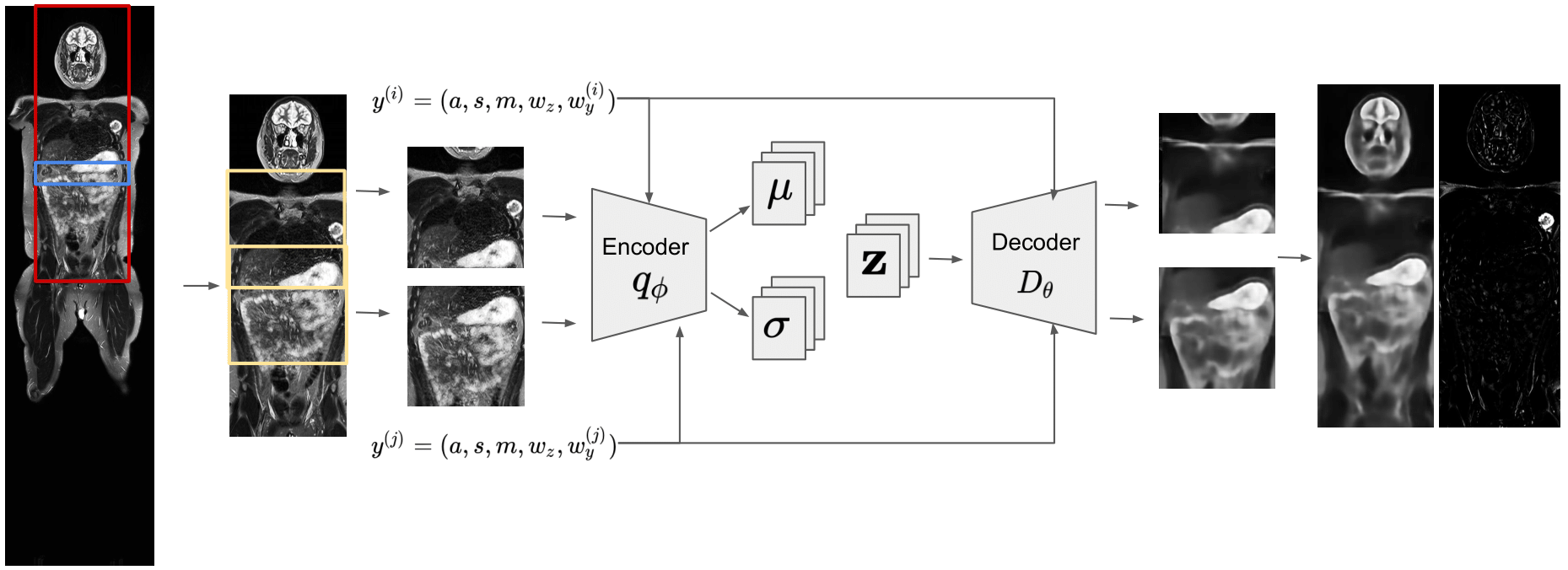}
\caption{Framework overview. The blue bounding box is determined by multi-scale template matching and determines the width of the chest. The red box then defines the region of interest. Square windows (yellow) within the region of interest are inputted to the spatial VAE.} \label{fig:overview}

\end{figure}

\section{Method}

\subsection{Conditional $\beta$-Variational Auto-Encoder} 

\noindent
\textbf{Architecture:} We utilize a VAE to learn the distribution of cancer-free images such that the model fails to reconstruct anomalous regions (i.e. the cancerous part) of the test images. Note that while we build on \cite{baur2018}, we simplified this model by omitting the adversary as it has been demonstrated to only marginally improve performance. We input conditional variables to the encoder and decoder networks by using a linear layer to predict a scaling and bias factor for the batch normalization feature maps in each residual block. Unlike BigGAN, we do not include the latent samples in this linear layer due to the large dimensionality of the spatial latent space. \newline

\noindent
\textbf{Training} For image $\mathbf{x} \in \mathbb{R}^{c \times 256\times 256}$ where $c$ is the number of channels/slices, and conditional variable $y = ($age, sex, weight$, w_z, w_y)$ with slice height $w_z$ and window height $w_y$, the VAE parametrizes latent distribution $q_{\phi}(\mathbf{z}|\mathbf{x}, y)$ from which latent vector $\mathbf{\hat{z}}\sim q_{\phi}(\mathbf{z}|\mathbf{x}, y)$ is sampled via  the reparametrization trick. Reconstruction $\mathbf{\hat{x}} = D_\theta(\mathbf{\hat{z}}, y)$ is then generated with decoder $D_\theta$. We minimize the conditional $\beta$-VAE loss function

\[
||\mathbf{x} - \mathbf{\hat{x}}||^2 + \beta D_{KL}(q_{\phi}(\mathbf{z}|\mathbf{x}, y) || \mathcal{N}(0, \mathbf{I}))
\]

Intuitively, $\beta$ hyperparameter represents the strength on the KL Divergence regularization term which encourages the proposal distribution $q_{\phi}(\mathbf{z}|\mathbf{x}, y)$ to match a standard Normal prior. This constraint ensures that the latent space maintains a degree of smoothness and prevents the model from memorizing the training set as in the case of autoencoders. \newline

\begin{table}[t]
\caption{
Processing volumes of wbMRI scans as multiple slices instead of individual images (1 slice) achieves significantly better results. Also, it is crucial to provide the coordinates of the windows ($w_y, w_z$ ) in our sliding window approach. }\label{tab1}
\centering
\begin{tabular}{p{3cm} p{3cm} >{\centering\arraybackslash}p{1.5cm} >{\centering\arraybackslash}p{1.5cm} >{\centering\arraybackslash}p{1.5cm}}
\toprule
Dataset & Input & AUPRC & AUROC & DICE \\

\midrule
\multirow{2}{*}{\shortstack[l]{Chest w/ \\ simulated nodules}} & 1 slice & 0.221 & 0.912 & 0.251  \\
& 1 slice + $w_z$ & \textbf{0.283} & \textbf{0.941} & \textbf{0.298} \\
\midrule
\multirow{7}{*}{\shortstack[l]{Upper body w/ \\ real nodules}} & 1 slice & 0.033 & 0.845 & 0.053  \\
& 1 slice + $w_z$ & 0.124 & 0.844 & 0.141 \\
& 1 slice + $w_y$ + $w_z$ & 0.129 & \textbf{0.910} & 0.159\\
& 3 slices + $w_y$ + $w_z$ & 0.141 & 0.901 & 0.161 \\
& 5 slices + $w_y$ + $w_z$ & 0.174 & 0.891 & 0.189 \\
& 7 slices + $w_y$ + $w_z$ & 0.169 & 0.869 & 0.178 \\
& 9 slices + $w_y$ + $w_z$ & \textbf{0.179} & 0.894 & \textbf{0.191} \\

\bottomrule
\end{tabular}
\end{table}

\noindent
\textbf{Anomaly Detection} To produce an anomaly mask that we report back to the user, we reconstruct $\mathbf{x}$ with maximum likelihood $\mathbf{z}$ from distribution $q_\phi(\mathbf{z}|\mathbf{x},y)$ ($\mathbb{E}[q_\phi(\mathbf{z}|\mathbf{x},y)]$) and obtain the pixel-wise squared-error loss mask. We then apply the post-processing steps by only keeping loss values for which the reconstruction underestimates the pixel value since lesion are brighter than normal tissues on STIR images and remove small regions which are unlikely to represent lesions using a $3 \times 5 \times 5$ median filter. During training, we randomly sample a window with a random height within the region of interest and during testing, we produce anomaly masks for equally spaced vertical windows and average overlapping mask regions on the same slice. 




\subsection{3D Context and Patient Features}
In prior work, images were treated independently which failed to capture the spatial correlation between slices when they are actually sampled from the same 3D volume. Meanwhile, training a 3D convolutional model is computationally expensive and requires a consistent shape for the input volume which is difficult to achieve in pediatric population. We therefore leverage information of neighbouring slices by using consecutive slices as a multi-channel input. The residuals of the middle channel, which we observe to have the lowest loss (and most accurate reconstruction), are then used to produce the loss mask. We slide the window across slices and concatenate loss masks before post-processing with the aforementioned method. Baur et al. demonstrate that a spatial latent space is essential for accurate reconstructions of brain MRI as it facilitates the comparison of similar structures \cite{baur2018}. Because we employ a sliding window approach, we condition the model on the window's coordinates since image contents greatly depends on the location of the window. We compute the slice coordinates using the slice spacing and thickness and use the relative distance from the back of the patient. All conditioned variables are then standardized.

\section{Experiments}

\subsection{Experimental Setup}
\textbf{Dataset:} We demonstrate the efficacy of our method on a dataset of coronal STIR whole-body MRI provided by a children's hospital containing 535  non-cancerous images (22 to 45 slices each for a total of 18012 slices) and 27 annotated images with at least one lesion. Images are preprocessed with histogram equalization, N4 bias field correction, and noise removal. Pediatric wbMRI are more difficult to register due to the greater variation in body positions. Because we also lack the annotations and atlas required for most robust registration methods, we focus on modeling a region of interest which includes the width of the chest from head to pelvis. We employ a pseudo-registration method by first localizing the bounding box which corresponds to each patient's chest with multi-scale template matching. We then define the region of interest as the chest width from top to middle pixel row of the image (Figure~\ref{fig:overview}). We proportionally resize each region of interest to a width of 256 pixels. Because the height of the regions of interest differs between images, we use a vertical sliding window approach to model one square $256 \times 256$ window at a time. \newline
\\
\noindent
\begin{table}[t]
\caption{Ablation on using various patient features (age, weight, sex). Here window coordinates are still provided and we process 5 MRI slices. Additional features improves performance substantially. However, $\beta$ needs to be tuned appropriately. }\label{tab2}
\centering
\begin{tabular}{
>{\centering\arraybackslash}p{1.2cm} 
>{\centering\arraybackslash}p{1.2cm} 
>{\centering\arraybackslash}p{1.2cm} 
>{\centering\arraybackslash}p{1.2cm} 
>{\centering\arraybackslash}p{1.2cm} 
>{\centering\arraybackslash}p{1.5cm} 
>{\centering\arraybackslash}p{1.5cm} 
>{\centering\arraybackslash}p{1.5cm}
}
\toprule
Slices(5)   & Age & Weight & Sex & Tune $\beta$ & AUPRC & AUROC & DICE\\
\midrule
\checkmark &     &       &     & & 0.174 & 0.891 & 0.189\\
\checkmark & \checkmark & & & & 0.169 & 0.853 & 0.184\\
\checkmark & & \checkmark &  & & 0.135 & 0.822 & 0.133\\
\checkmark & & & \checkmark & & 0.213 & 0.917 & 0.194\\
\checkmark & \checkmark & \checkmark & \checkmark &  & 0.171 & 0.894 & 0.196\\
\checkmark & & & \checkmark & \checkmark & \textbf{0.235} & \textbf{0.941} & \textbf{0.239}\\

\bottomrule
\end{tabular}
\end{table}

\noindent
\textbf{Metrics:}
We evaluate model performance with the Area Under the Precision-Recall Curve (AUPRC), Area Under the Receiving-Operator Curve (AUROC), and DICE coefficient. Both AUPRC and AUROC allow the evaluation of model capabilities without deciding an operating point or decision threshold. We note that the AUROC is sensitive towards the minority class in the context of severe class imbalance. To select a threshold required to compute DICE, we use the 99th percentile of non-zero loss values after post-processing. \newline

\noindent
\textbf{Labeling:} We conduct two sets of anomaly detection experiments, one on images with synthetically generated tumors and one on real tumors labelled by a radiologist. We generate the synthetic tumors by adding a Gaussian blob with peak intensity from 0.6 to 3 consecutive slices of each test volume. \newline

\noindent
\textbf{Implementation details:} 
All models are trained with a 0.001 learning rate, 16 batch size, and Adam optimizer for 150 epochs and the spatial latent dimension is maintained at $32 \times 16 \times 16$. We also anneal $\beta$ during training to linearly increase during the first 20 epochs. We modify the encoder and decoder networks from a series of ReLU activated convolutional layers to a sequence of residual blocks which is used in the state-of-the-art conditional GAN model (BigGAN \cite{brock2018large}). All model training is done in PyTorch v1.6.0 and required approximately 24 hours on 2 NVIDIA T4 GPUs. 
\begin{figure}[t]
    \centering
    \includegraphics[width=0.98 \linewidth]{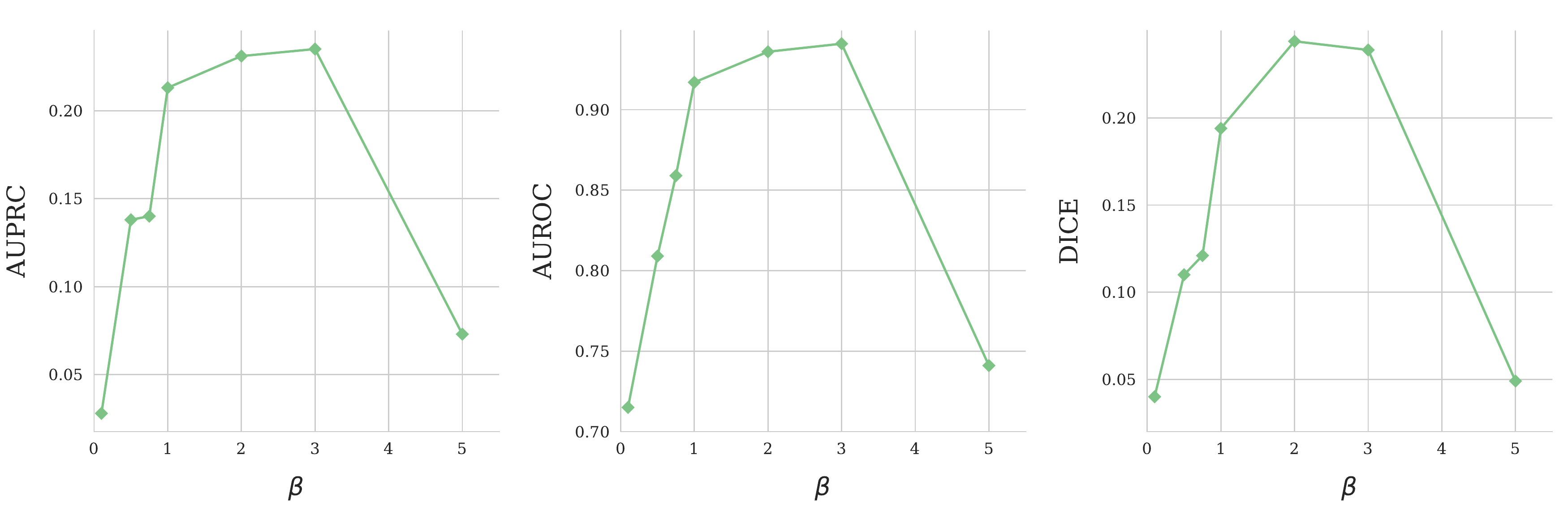}
    \caption{Performance comparison for various settings of the regularization hyperparameter, $\beta$, with 5 slice inputs and conditioning on patient sex.}
    \label{fig:beta}
\end{figure}


\subsection{Cancer Detection}
\noindent
\textbf{3D Context:} To answer whether the naive baseline approach benefits from conditioning on the slice coordinate, we compare the methods using only crops of the chest nodules in the test set (Table \ref{tab1}). We observe a 0.062 increase in AUPRC performance and 0.047 increase in DICE. We then proceed to model the larger region of interest using the sliding window approach. With the naive independent window samples, we observe that the AUPRC and DICE scores are low while the AUROC is relatively high compared to the value of a random classifier. Upon only adding the slice location, the model considerably improves (0.091 AUPRC and 0.088 DICE increases) while subsequently adding the window height marginally improves AUPRC and DICE. Maintaining both window and slice coordinates while increasing the number of slices/channels further boosts performance up to 5 slices (0.017 AUPRC improvement over 1 slice) where it starts to plateau. We therefore used 5 slices for the remaining experiments to reduce runtime. This suggests that the 3D information from neighbouring slices helps learn a more discriminative latent manifold. \newline


\noindent
\textbf{Patient Features:} Next, we experimented with the inclusion of patient's age, weight, and sex as these meta-features theoretically correlate with the anatomy. In these experiments window coordinates are still provided and we process 5 MRI slices to minimize excess computation. We observe that the model benefits from the inclusion of the patient's sex (0.061 AUPRC improvement) while including age and weight worsen the performance. We hypothesize that this is due to the fact that sex is a binary attribute which helps to distinguish between real differences on the wbMRIs while substantial number of training examples still exists for both groups (sexes). In contrast, the continuous attributes of weight and age may be poorly represented for some ranges, giving a greater chance of introducing biases with our relatively small training set. \\

\begin{figure}[t]
\includegraphics[width=0.75\linewidth]{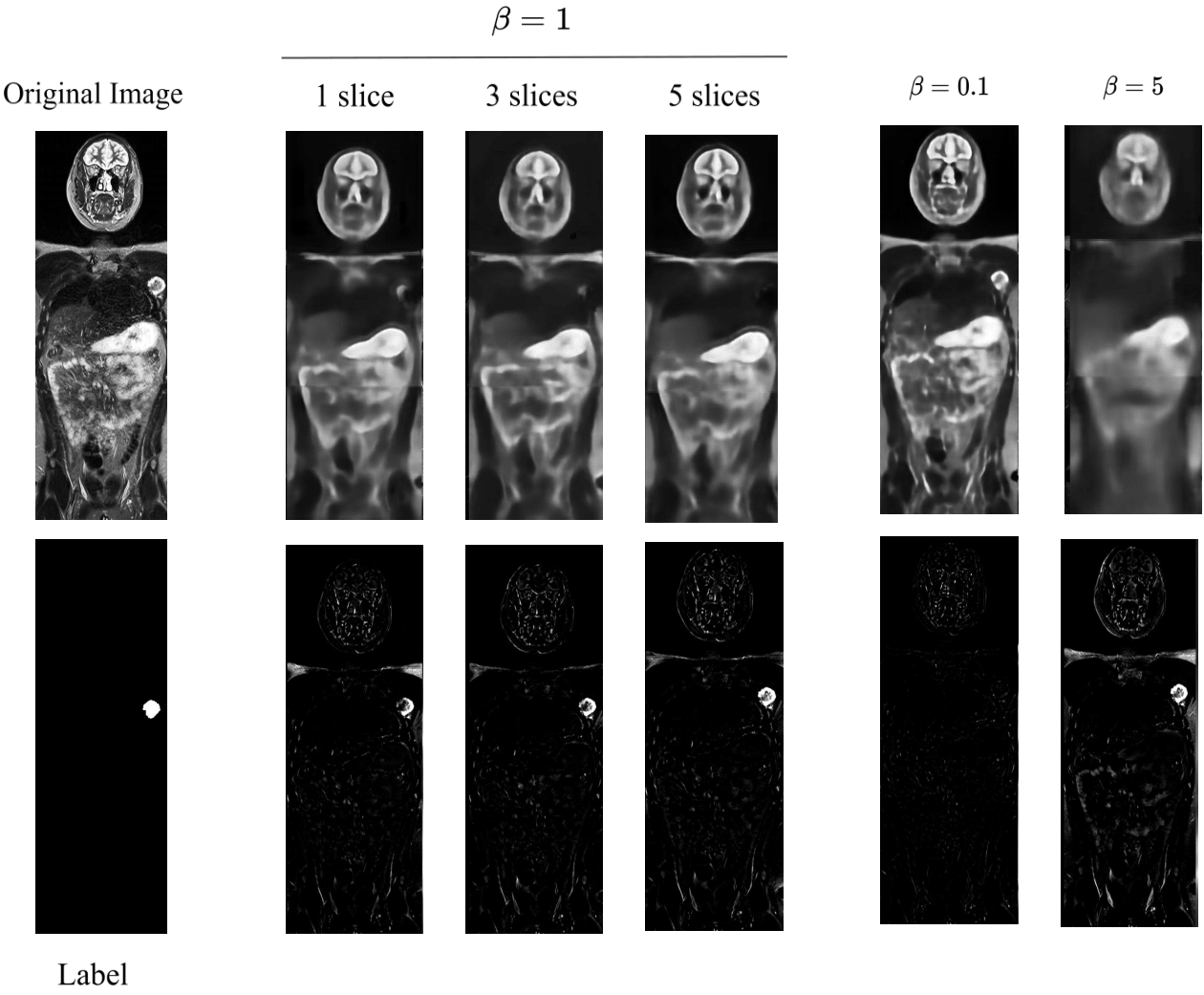}
\centering
\caption{Stitched reconstructions (top row) and their resulting postprocessed loss masks (bottom row) for the original image with ground-truth label in the leftmost column. We observe a more accurate loss mask with the increase in number of slices used for input. Moreover, we observe blurrier reconstructions, capturing more global features with the increase of $\beta$.} \label{fig:wbmri_masks}
\end{figure} 
 
\noindent
\textbf{Regularization:} Table \ref{fig:beta} demonstrates that $\beta = 2$ and $\beta = 3$ provide the greatest AUPRC and DICE respectively, while too low or too high $\beta$ values decrease the performance. Figure~\ref{fig:wbmri_masks} shows that with a low $\beta$ value the reconstructions are very close to the input but the model fails to learn the healthy latent manifold which results in reconstructions of anomalies. As $\beta$ increases, the cancerous image region ceases to be reconstructed, but as $\beta$ gets too high the input's details are lost which contributes to more false positives.


\section{Conclusion}

We demonstrated applicability of unsupervised cancer detection from wbMRI in pediatric population. Using a multi-channel sliding window approach, we incorporated 3D information which overcomes several challenges specific to this modality such as registration and high image dimensionality and show improved performance over a 2D baseline. We show that conditioning on certain meta-features which are indicative of window contents improves detection performance and that tuning the regularization hyperparameter can further help disentangle the latent space.
%
%
%

\newpage
\bibliographystyle{splncs04}
\bibliography{references.bib}

%
\end{document}